\documentclass[nofootinbib,twocolumn,aps,pre,superscriptaddress,citeautoscript,floatfix]{revtex4-2}
\usepackage{graphics}
\usepackage[title]{appendix}
\usepackage{graphicx}
\usepackage{color}
\usepackage{bm}
\usepackage{amsmath}
\usepackage{amssymb}
\usepackage{float}

\begin{document}

\title{Universality in the dynamics of vesicle translocation through a hole}

\author{Bin Zheng}
\affiliation{Wenzhou Institute, University of Chinese Academy of Sciences, Wenzhou, Zhejiang 325001, China}
\affiliation{Oujiang Laboratory, Wenzhou, Zhejiang 325000, China}

\author{Fangfu Ye}
\affiliation{Wenzhou Institute, University of Chinese Academy of Sciences, Wenzhou, Zhejiang 325001, China}
\affiliation{Oujiang Laboratory, Wenzhou, Zhejiang 325000, China}
\affiliation{School of Physical Science, University of Chinese Academy of Sciences, Beijing 100049, China}
\affiliation{Beijing National Laboratory for Condensed Matter Physics, Institute of Physics, Chinese Academy of Sciences, Beijing 100190, China}

\author{Shigeyuki Komura}
\affiliation{Wenzhou Institute, University of Chinese Academy of Sciences, Wenzhou, Zhejiang 325001, China}
\affiliation{Oujiang Laboratory, Wenzhou, Zhejiang 325000, China}
\affiliation{Department of Chemistry, Graduate School of Science, Tokyo Metropolitan University, Tokyo 192-0397, Japan}

\author{Masao Doi}
\affiliation{Wenzhou Institute, University of Chinese Academy of Sciences, Wenzhou, Zhejiang 325001, China}
\affiliation{Oujiang Laboratory, Wenzhou, Zhejiang 325000, China}

%\date{\today}

\begin{abstract}
We analyze the translocation process of a spherical vesicle,  made of membrane and 
incompressible fluid, through a hole smaller than the vesicle size,
driven by pressure difference $\Delta P$.  
We show that  such a vesicle shows certain universal characteristics which is independent of the details 
of the membrane elasticity;  (i) there is a  critical pressure 
$\Delta P_{\rm c}$ below which no translocation occurs, (ii) $\Delta P_{\rm c}$  decreases to zero
as the vesicle radius $R_0$ approaches the hole radius $a$, satisfying the scaling relation  
$\Delta P_{\rm c} \sim (R_0 - a)^{3/2}$,
and (iii) the translocation  time $\tau$ diverges as 
$\Delta P$ decreases to $\Delta P_{\rm c}$, satisfying the scaling relation  
$\tau \sim (\Delta P -\Delta P_{\rm c})^{-1/2}$.  
\end{abstract}

\maketitle

\section{Introduction}
A vesicle is a fluid droplet covered by a membrane. 
Examples of natural vesicles are cells and organoids.
Artificial vesicles made of amphiphilic copolymers, called polymersomes, have been developed and  
are used as a capsule of drugs or RNA fragments~\cite{Lee2012,Rideau2018,Bohdana1999,Dennis2002,Carla2014}. 
Translocation of such objects through a constriction  is ubiquitous in many biological and 
bio-technical systems, such as  blood flow, cell circulation, drug delivery, and cell 
manufacturing~\cite{Guo2003,Adrian2017,Zhang2018, Plaks2013, Zhifeng, Colin2017, Vormittag2018}.  

Many theoretical studies and simulations have been done for vesicle 
translocation~\cite{Gompper1995,Linke2006, Tordeux2002,Bogdan2022, Meng2014, Han2019}, but our understanding for these phenomena is still far from complete. 
One reason is that there exist many types of membranes (lipid membranes, 
polymer membranes, cell membranes), and it is difficult to construct a general theory valid 
for all vesicles.  Indeed, various models have been used in the past.
In the early works~\cite{Gompper1995, Linke2006}, 
it was assumed that the vesicles have constant surface area but have  
variable volume (as the fluid can permeate through the membrane).  
In recent works, alternative models have been used in which
the vesicle has a constant volume, but has a variable surface 
area~\cite{Tordeux2002, Hamid2016, Petch2018}.   Another important difference is
whether the membrane is a fluid which has zero shear modulus in the plane, or an elastic sheet which has non-zero shear modulus~\cite{Nelson}. 
In most of the previous works, the membrane has been assumed to be 
a fluid membrane, but an elastic membrane, often called the tethered membrane, exists and is important in many biological systems. Translocation of such membranes has been studied 
recently by computer simulations~\cite{Bogdan2022}. 

In this work, we develop a theory for vesicle translocation using a model which can
 include a large class of membranes such as fluid membranes, tethered membranes, 
 and  composite membranes.  The model assumes that 
(i) the elastic energy of the membrane is due to the in-plane 
stretching of the membrane (i.e., the bending energy of the membrane
is ignorable), (ii) the volume of the vesicle is kept constant during translocation (i.e., the fluid 
permeation through the membrane is  ignorable), and (iii) the vesicle takes a spherical 
shape outside of the hole.   We show that if a vesicle satisfies these conditions, 
its translocation behavior  has certain universal characteristics which are independent 
of the details of the membrane property. 

The outline of the article is as follows. In Sec.~II, we discuss the general form of the free energy 
of a stretchable membrane model, and give two specific models: (i) fluid membrane model and (ii) rubber membrane model. Time-evolution equation for the vesicle during translocation is derived. 
In Sec.~III, we show the free energy profile for the translocation process, and  calculate the
critical pressure needed to cause the translocation.  We then show that there is certain universality 
in the critical pressure and the translocation time. Finally, in Sec.~IV, we summarize our result 
and compare it with previous simulation works.

%%%%%%%%%%%%%%%%%%%%%%%%%%%
\begin{figure*}[tbh]
\centering
\includegraphics[width=0.8\textwidth,draft=false]{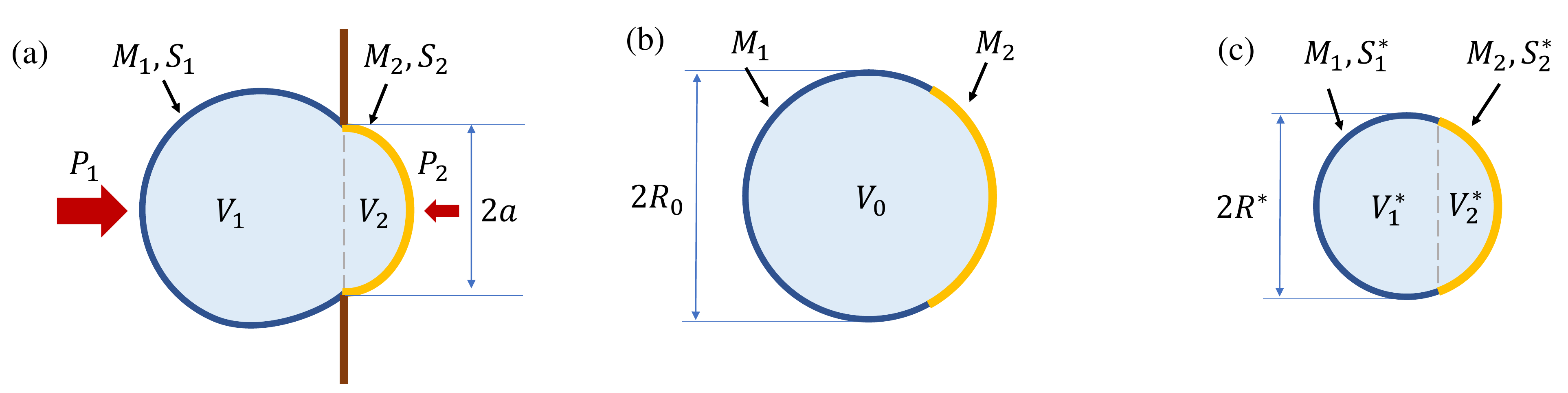}
\caption{ (a) When a vesicle is translocating through a hole of radius $a$,  it 
 takes a double-spherical cap shape.  The volume of the vesicle in each chamber is 
$V_1$ and $V_2$, and the area and the mass of the membrane in each chamber are $S_i$ 
and $M_i$ ($i=1,2)$, respectively.    (b) Outside of the hole, the vesicle takes a spherical shape of 
volume $V_0=V_1 + V_2$. The parts of the membrane which are in chamber 1 or 2 
in the state (a) are shown by different colors.  (c) The vesicle in the reference state where
the vesicle takes a spherical shape of radius $R^{*}$.  The stretching $\lambda_i$ 
of the membrane is defined by the length change with respect to this state.  
Notice that  the volume $V_0$ in the state (b) is generally different
from the volume $V^*=V_1^* +V_2^*$ in the state (c).}
\label{fig1}
\end{figure*}
%%%%%%%%%%%%%%%%%%%%%%%%%%%%

\section{Theory}
\subsection{General membrane model}
We consider a spherical vesicle consisting of membrane and inner fluid.  
The membrane can be a single layer of amphiphilic molecules
(uncross-linked or cross-linked polymers as in polymersomes) or composite membranes 
made of a fluid layer and elastic networks (as in plasma membranes). 

We regard the membrane as a 2D surface, described by a 
3D position vector $\mathbf{r}(s_1, s_2)$,  where $s_1$ and $s_2$ are the 2D coordinates
which are defined in the reference state and identify a point in the membrane.   
We assume that, in the reference state, $\partial  \mathbf{r}/\partial s_1$ 
and $\partial  \mathbf{r}/\partial s_2$ are unit vectors orthogonal to each other.  
(When we use the spherical coordinate $(\theta, \psi)$, $s_1$ and $s_2$ can be written  
as $s_1=R^* \theta$ and  $s_2=(R^* \sin \theta) \psi$, where $R^*$ is the radius of the sphere.) 
We assume that the membrane is homogeneous 
and isotropic in the tangential plane. We also assume that the membrane is thin 
and ignore the bending energy. The elastic energy of the membrane is therefore 
due to the in-plane membrane stretching and can be characterized by the 
function $f(\lambda_1, \lambda_2)$, where $\lambda_1$ and $\lambda_2$ are 
the principal values of stretching and $f(\lambda_1, \lambda_2)$ is the 
elastic energy per unit area in the reference state.   ($\lambda_1$ and $\lambda_2$ 
are given by the square root of the eigenvalues of the $2 \times 2 $ matrix 
$B_{ij}= (\partial \mathbf{r}/\partial s_i )\cdot  (\partial \mathbf{r}/\partial s_j) $ 
with $i,j=1,2$.)  The total elastic energy functional of the membrane is given by 
the integral  of $f(\lambda_1, \lambda_2)$ over the entire surface:
\begin{equation}
\mathcal{A}[\mathbf{r}(s_1, s_2)] =\int ds_1 ds_2\, f(\lambda_1, \lambda_2).    
\label{eqn:1a0}
\end{equation}
The equilibrium state of the membrane is calculated by minimizing this functional with 
respect to  $\mathbf{r}(s_1, s_2)$.

For fluid membranes, $f(\lambda_1, \lambda_2)$ depends only on the 
area change ratio $\lambda_1\lambda_2$.  In this work, we consider the following model energy
\begin{equation}
f_{\rm FM}(\lambda_1, \lambda_2)= \frac{k_{\rm a}}{2} (\lambda_1 \lambda_2 -1)^2,
\label{eqn:7a}
\end{equation}
where $k_{\rm a}$ is the elastic constant, which has the same dimension as the surface tension.

For elastic (or tethered) membranes,  $f$ cannot be written in this form because 
they experience restoring force caused by shear strain $\lambda_1 - \lambda_2$.
As an example,we consider the following Neo-Hookean model~\cite{DoiSoftMatterPhysics},
which has been extensively used as a model of rubber membrane  (such as balloons)  
\begin{equation}
f_{\rm RM}(\lambda_1, \lambda_2) 
= \frac{\mu d}{2} [ \lambda_1^2+  \lambda_2^2 + (\lambda_1 \lambda_2)^{-2} -3 ],
\label{eqn:1a3}
\end{equation}
where $\mu$ is the 3D shear modulus of the rubber and $d$ is the thickness of the membrane.
We call such a membrane a rubber membrane.

\subsection{Free energy of a vesicle under translocation}
We now consider the situation that such a vesicle is forced to pass through a hole of radius $a$ 
made in a wall  at the boundary between two chambers 1 and 2, as shown in Fig.~\ref{fig1}. 
Initially, the vesicle has a spherical shape of radius  $R^\ast$ [Fig.~\ref{fig1}(c)]
and it is inflated to have a radius $R_0$ by some methods (e.g., osmotic swelling or injection) [Fig.~\ref{fig1}(b)].   We use the swelling ratio defined by $\eta=R_0/R^\ast \ge 1$.
The pressure in each chamber is $P_1$ and $P_2$, and the vesicle is pushed by the  pressure 
difference $\Delta P=P_1-P_2>0$.   During the translocation, the vesicle takes a double spherical-cap 
shape, each having volume $V_1$ and $V_2$ [Fig.~\ref{fig1}(a)].  
Since the membrane is assumed to be impermeable to the fluid during the passage, the total fluid 
volume $V_{0}=V_1+V_2 =4\pi R_{0}^3/3$ remains unchanged.
Let $S_1$ and $S_2$ be the area of the membrane in each
chamber.  Since the membrane can be stretched, the total area of the membrane 
$S=S_1+S_2$ can vary.
To describe the transport of the membrane, we introduce the membrane mass $M_1$ and $M_2$
in each chamber.  
The total  mass $M_{0}=M_1+M_2$ remains constant, but the membrane mass density 
$M_i/S_i$ ($i=1,2$) in each chamber  changes in time. 

Given the free energy functional such as Eq.~(\ref{eqn:1a0}),  the 
free energy of the vesicle is uniquely determined as a function of $M_i$ and $V_i$, 
and is written as
\begin{equation}
G=A(M_1, V_1)+A(M_2, V_2)+P_1V_1+P_2V_2, 
\label{eqn:1}
\end{equation}
where $A(M_i, V_i)$ is the minimum of the functional $\mathcal{A}[\mathbf{r}(s_1,s_2)]$ 
subject to the constraint that the membrane has mass $M_i$ and includes fluid volume $V_i$, 
and that its circular edge is fixed to the hole of radius $a$.  In the following, we calculate $A(M_i, V_i)$ explicitly for a fluid membrane and a rubber membrane. 

\subsubsection{Fluid membrane}
We first consider the fluid membrane model. The fluid membrane at equilibrium
always takes a spherical-cap shape with constant curvature.  
Therefore the integral of Eq.~(\ref{eqn:7a}) can be written as $S^\ast_i f(S_i/S^\ast_i)$, and the total free energy can be written as 
\begin{align}
	G(M_i, V_i)&=\frac{k_{\rm a}}{2}\frac{[S_1(V_1)-S_{\rm 1}^*(M_1)]^2}{S_{\rm 1}^*(M_1)} \nonumber \\ 
	&+\frac{k_{\rm a}}{2}\frac{[S_2(V_2)-S_{\rm 2}^*(M_2)]^2}{S_{\rm 2}^*(M_2)}+P_1 V_1+P_2 V_2, 
	\label{energy1again}
\end{align}
where $S^\ast_i =4 \pi (R^\ast)^2 M_i/M_0$
is the area of the membrane of mass $M_i$ in the reference state [Fig.~\ref{fig1}(c)].
The surface area $S_i$ is determined by the condition that the spherical cap has volume $V_i$ 
and its edge is fixed to a circle of radius $a$. 
This condition leads to $S_i = \pi (a^2 + h_i^2)$,  where $h_i$ is the solution of the 
equation $V_i =  \pi h_i(3 a^2 + h_i^2)/6$ for given $V_i$. 
Therefore, the energy of the fluid membrane can be  
expressed in terms of $M_i$ and $V_i$.

Since $V_{0}=V_1+V_2$ and  $M_{0}=M_1+M_2$ are constant, the total free energy 
$G$ can be expressed as a function of $V_2$ and $M_2$.  
We define the following dimensionless quantities
\begin{equation}
	x= \frac{M_2}{M_0} - \frac{1}{2} , \qquad
	y= \frac{V_2}{V_0} - \frac{1}{2},        
	\label{eqn:5}
\end{equation}
and write the total free energy as
\begin{align}
	G(x,y)&=\frac{k_{\rm a}}{2}\frac{[S_1(y)-(1/2-x)S^\ast]^2}{(1/2-x)S^\ast} \nonumber \\ 
	&+\frac{k_{\rm a}}{2}\frac{[S_2(y)-(1/2+x)S^\ast]^2}{(1/2+x)S^\ast}-\Delta P V_0 y,
	\label{energy1againdim}
\end{align}
where $S^*=S_0/\eta^2$ and $\eta=R_0/R^*$ is the swelling ratio. In addition, $S_2/S_0$ can be derived from the geometric relation for a spherical cap as mentioned above, and becomes
\begin{align}
\label{eqgeometric1} \frac{S_2(y,a/R_0)}{S_0}=-\frac{(a/R_0)^2}{4}+\frac{(a/R_0)^4}{4\beta^{2/3}}+\frac{\beta^{2/3}}{4}, 
\end{align}
where 
\begin{align}
\label{eqgeometric2}  \beta(y,a/R_0)=4(y+1/2)+\sqrt{(a/R_0)^6+16(y+1/2)^2}.
\end{align}
Similarly,  $S_1(y, a/R_0)/S_0$ is obtained by replacing $y$ with $-y$ in the above equation. 

\subsubsection{Rubber membrane}
%%%%%%%%%%%%%%%%%%%%%%%%%%%%%
We next consider the rubber membrane whose free energy is given by Eq.~(3).
To calculate the elastic free energy, we first focus on the spherical cap in chamber 2, 
which has the surface mass $M_2$ and is initially a part of the equilibrium vesicle of radius $R^*$.
Then, it is fixed to the hole of radius $a$ and is inflated to the volume $V_2$ as shown by the yellow curve in Fig.~\ref{fig1}(a).  

\begin{figure}[tbh]
	\centering
	{\includegraphics[width=0.4\textwidth,draft=false]{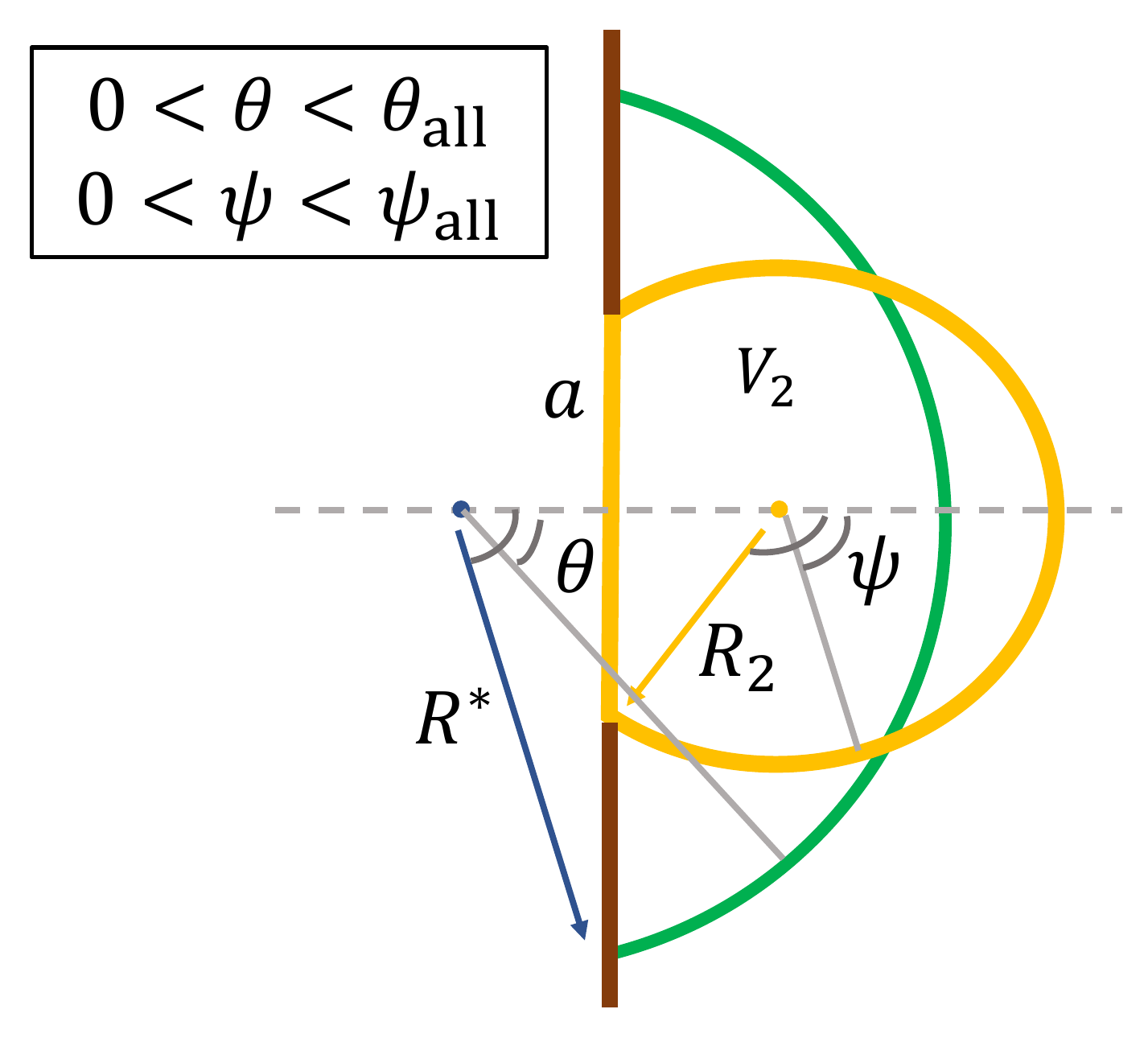}}
	\caption{
		Schematic illustration of the rubber membrane model. 
	}
	\label{fig2}
\end{figure}

Exact calculation of the elastic energy for such deformation is 
difficult for a rubber membrane.  Here we use an approximation
and assume that the membrane takes a spherical shape of 
curvature $1/R_2$, as shown in Fig.~\ref{fig2}.  Then, a point located at a polar angle $\theta$ on the sphere
of radius $R^*$ in the reference state will move to a point located at a polar angle
$\psi(\theta)$ on the inflated sphere of radius $R_2$. 
This deformation elongates the line segment along the $\theta$ direction 
by factor $R_2 d\psi/R^* d \theta$, and the line segment perpendicular to this
direction by factor $R_2 \sin \psi/R^* \sin \theta$.  
Hence the principal values of stretching in chamber 2 are given by 
\begin{align}
	\lambda_{21}=\frac{R_2}{R^*} \frac{d\psi}{d\theta}, \qquad 
	\lambda_{22}=\frac{R_2 \sin \psi}{R^* \sin \theta}.
\end{align}

To determine the functional form of $\psi(\theta)$, we assume that the surface area 
of the membrane changes uniformly by deformation, such that $\lambda_{21} \lambda_{22} $ is 
constant (independent of $\theta$).  This condition gives
\begin{equation}
		 \frac{ \sin \psi}{\sin \theta}\frac{d\psi}{d\theta} = {\rm Const.}
\end{equation}
From this relation, we have 
\begin{equation}
	\frac{ 1- \cos {\theta}}{ 1- \cos {\psi}}= \frac{ 1- \cos {\theta_{\rm all}}}{ 1- \cos {\psi_{\rm all}}},
\end{equation}
where $\theta_{\rm all}$ and $\psi_{\rm all}$ are the maximum values of $\theta$ and $\psi$, respectively, which
are given by (see Fig. \ref{fig2})
\begin{align}
	& \theta_{\rm all}=\arccos(-2x), \\
	& \psi_{\rm all}=\arccos\left[1-\frac{2S_2/S_0}{(R_2/R_0)^2}\right].
\end{align}
Here, $R_2(y)/R_0$ can be derived from the geometric relation for a spherical cap, $S_2=2\pi R_2 h_2$, 
where $h_2$ is the solution of the equation $V_2=\pi h_2^2(3R_2-h_2)/3$. On the other hand, $S_2(y,a/R_0)/{S_0}$ is given by Eqs.~(\ref{eqgeometric1}) and (\ref{eqgeometric2}). Then, the same calculation can be done for the membrane in chamber 1.

The total energy of deformation can be expressed by the integrals over $\theta$ 
\begin{align}
	G(x,y) =&\frac{\mu d}{2}\int^{\pi-\theta_{\rm all}}_{0} d \theta \, \left(\lambda_{11}^2+\lambda_{12}^2+\frac{1}{\lambda_{11}^2 \lambda_{12}^2}-3\right) \nonumber \\
	&\times 2\pi (R^*)^2 \sin \theta \nonumber \\
	+&\frac{\mu d}{2}\int^{\theta_{\rm all}}_{0} d \theta \, \left(\lambda_{21}^2+\lambda_{22}^2+\frac{1}{\lambda_{21}^2 \lambda_{22}^2}-3\right)  \nonumber \\ 
	&\times2\pi (R^*)^2\sin \theta -\Delta P V_0 y.
\label{eqn:nh}
\end{align}
Although, the above integrals can be performed analytically, it is too lengthy to present the result here. 

\subsection{Time-evolution equation}

If the free energy of the system is expressed as a function of $x$ and $y$,
their time evolution can be discussed in the same way as in the previous works~\cite{Petch2018}.
Here, it is important to note that the characteristic times of $x$ and $y$
are quite different from each other.  The membrane mass transport is governed 
by the solid friction of the membrane sliding against the 
hole, while the fluid volume transport is governed by the fluid flow through the hole.   
Since the fluid friction is much smaller than the solid friction, the relaxation time of $y$ is
much smaller than that of $x$.  Therefore we may assume that 
$y$ is at equilibrium for a given value of $x$. 
Hence, $y$ is determined by the condition
\begin{equation}     
	\frac{\partial G(x,y)}{ \partial y} =0.        	
	\label{eqn:6}
\end{equation}
Let $y=y^*(x)$ be the solution of this equation. 
The total free energy is therefore written as a function of $x$ only and we write it as
\begin{equation}
 G^*_x = G(x, y^*(x)).
\end{equation}   
Then the time-evolution equation for $x$ is written as
\begin{equation}
	\xi(x) \frac{dx}{dt} = - \frac{d G^*_x}{d x},     	
	\label{eqn:7}
\end{equation}
where $\xi(x)$ is the friction coefficient representing the solid friction for the membrane sliding at the hole rim.

\section{Results}
\subsection{Free energy profile}
%%%%%%%%%%%%%%%%%%%%%%%%%%%
\begin{figure}[tbh]
\centering
\includegraphics[width=0.45\textwidth,draft=false]{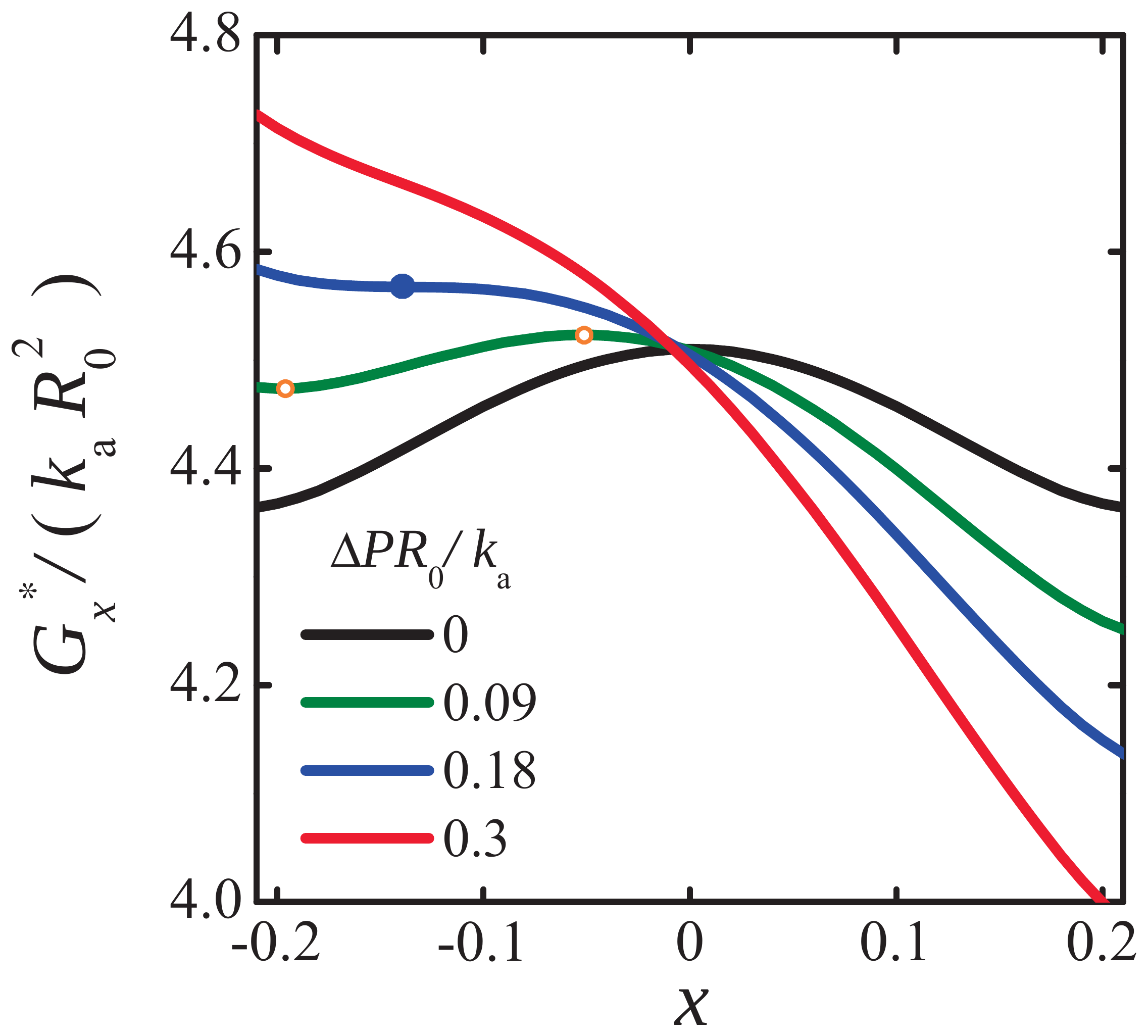}
\caption{
The free energy $G^*_x$ of a fluid membrane 
 as a function of $x=M_2/M_0 - 1/2$ when     
$a/R_{0}=0.9$ for different values of $\Delta P$.}
\label{fig3}
\end{figure}
%%%%%%%%%%%%%%%%%%%%%%%%%%%

Figure~\ref{fig3} shows an example of $G^*_x$  calculated for a fluid 
membrane whose free energy is given by Eq.~(\ref{energy1againdim}).
When there is no pressure difference, i.e., when $\Delta P=0$, $G^*_x$ has a peak at $x=0$.
When a small pressure is applied, $G^*_x$ has a local minimum in the region of $x<0$. 
This corresponds to the  situation that the vesicle is trapped by the hole. 
By the further increase of $\Delta P$, the local minimum becomes unstable
(see the curve of  $\Delta P R_0/ k_{\rm a} \approx  0.18$ in 
Fig.~\ref{fig3}).  The vesicle translocates  from chamber 1 to 2 
above this critical pressure difference.

\subsection{Critical pressure for translocation}
The above argument shows that the critical pressure difference 
$\Delta P_{\rm c}$  is determined by the following condition for general membranes
\begin{equation}
\frac{d G^*_x}{ d x }=\frac{d^2 G^*_x}{d x^2 }=0.     	
\label{eqn:8}
\end{equation}
Although $\Delta P_{\rm c}$ can be obtained by solving Eq.~(\ref{eqn:8}), 
the calculation becomes somewhat cumbersome. However, there is a simpler way of 
calculating $\Delta P_{\rm c}$.

We consider the function $G^*_y = G(x^*(y), y)$, 
where $x^*(y)$ is the solution of $\partial G(x,y)/ \partial x =0$.
In Appendix A, we show that the critical state can also be obtained by solving the 
following equations:
\begin{equation}
\frac{d G^*_y}{d y }=\frac{d^2 G^*_y}{d y^2}=0.     	
\label{eqn:9}
\end{equation}
The advantage of using $G^*_y$ is that $x^*(y)$ is independent of $\Delta P$ and 
satisfies the symmetry relation $x^*(-y)=-x^*(y)$.  
Hence, $G^*_y$ can be written in the following form:
\begin{equation}
G^*_y= A_{\rm tot}(y) - \Delta P V_0 y,     
\label{eqn:11}
\end{equation}
where $A_{\rm tot}(y)$ stands for the minimum of the total membrane elastic energy [i.e.,  $A_{\rm tot}(y)={\rm Min} (A(M_1,V_1)+A(M_2,V_2)$] subject to the constraint that the inner fluid volume in chamber 2 is 
$(1/2+y)V_0$. Notice that $A_{\rm tot}(y)$ satisfies the relation $A_{\rm tot}(-y)=A_{\rm tot}(y)$.

Further simplification is possible for fluid membranes for which  
$A(M_i, V_i)$ can be written as $S^\ast_i f(S_i/S^\ast_i)$. 
Then the condition $\partial G(x,y)/ \partial x =0$ gives the relation 
$S_1/S^\ast_1 =S_2/S^\ast_2= S/S^\ast$, where $S=S_1+S_2$ is the total surface area of the vesicle and $S^* = S_0/\eta^2$. The free energy  $G^*_y$  can be written as
\begin{align}
	G^*_y & = S_1^*f(S_1/S_1^*)+S_2^*f(S_2/S_2^*) - \Delta P V_0 y \nonumber \\
	& = S^*f(S/S^*) - \Delta P V_0 y.
\label{energy2}
\end{align}
For given $a$ and $R_0$, $S$ depends only on $y$.
For Eq.~(\ref{energy2}), the first equation in Eq.~(\ref{eqn:9}) gives the following equilibrium pressure $\Delta P$ at state $y$, 
\begin{equation}
	\Delta P = \frac{f'}{V_0} \frac{d S}{d y}.
\label{eqn:14}
\end{equation}
The second equation in Eq.~(\ref{eqn:9}) gives the following equation for $y_{\rm c}$ at the critical state
\begin{equation}
\frac{f''}{S^\ast}  \left( \frac{d S}{d y} \right)^2    
+f' \frac{d^2 S}{d y^2} =0, 
\label{eqn:15}
\end{equation}
where $f'$ and $f''$ stand for the first and the second derivatives of the function $f$, respectively. In addition, $S/S_0$ is derived from the geometric relation for a spherical cap and is given by 
Eqs.~(\ref{eqgeometric1}) and (\ref{eqgeometric2}).
Finally, the analytical expression of $\Delta P$ in Eq.~(\ref{eqn:14}) can be obtained, but it is too lengthy to present it here.

\begin{figure}[tbh]
\centering
\includegraphics[width=0.45\textwidth,draft=false]{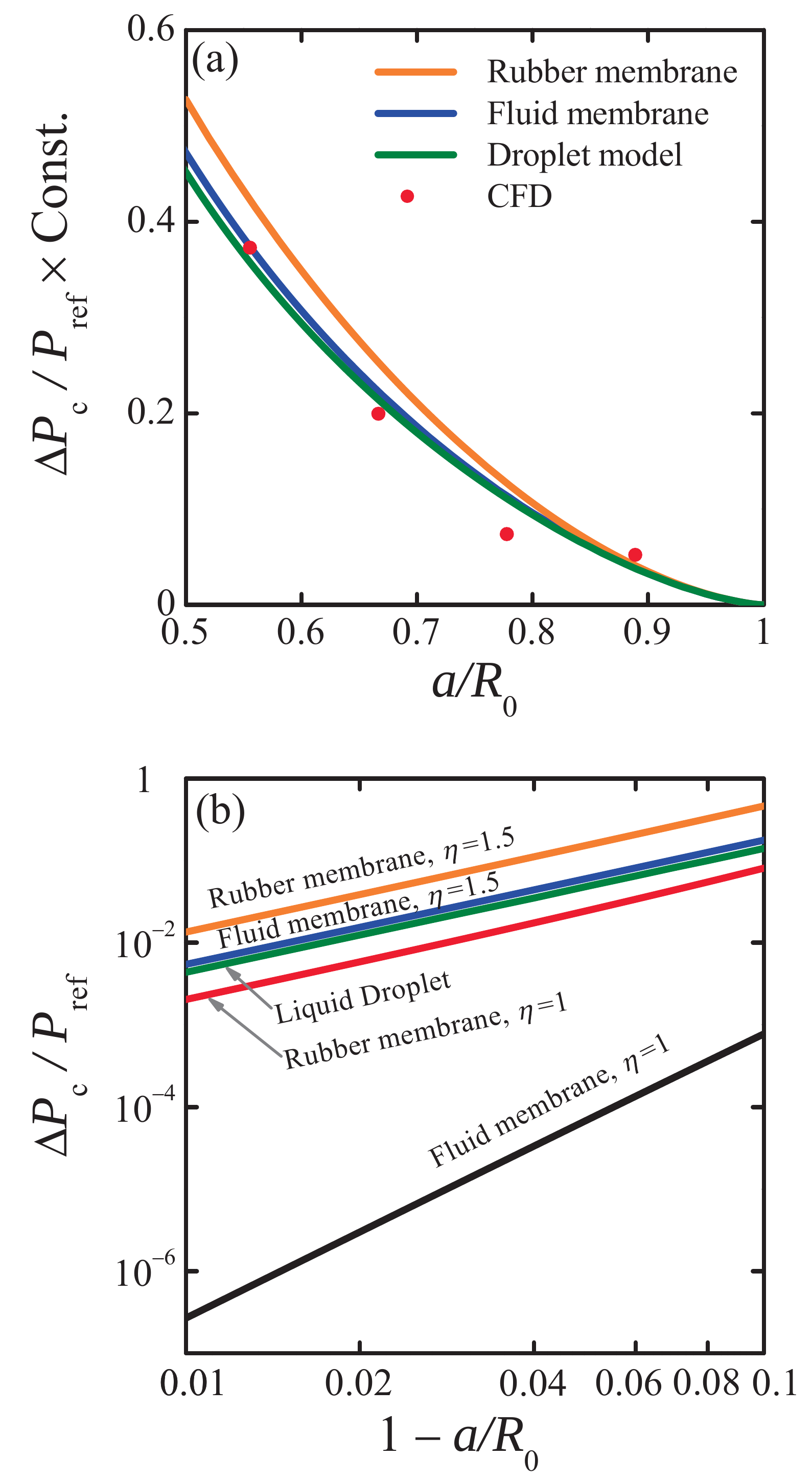}
\caption{(a) The reduced critical pressure $\Delta P_{\rm c}/P_{\rm ref}$ as a function of 
$a/R_0$ for different membrane models when $\eta=R_0/R^\ast=1.5$. 
Here, $P_{\rm ref}$ corresponds to $k_a/R_0$ for a fluid membrane, $\mu d/R_0$ for a rubber membrane, 
and $\gamma/R_0$ for a liquid droplet, respectively. 
The constant factors in $\Delta P_{\rm c}/P_{\rm ref}$ are chosen to have an overlap of 
different curves when $a/R_0$ is close to unity. 
The red circles are the data obtained by the computational fluid dynamics (CFD) in Ref.~\cite{Zhifeng}.
(b) The double logarithmic plot of (a) as a function of $1-a/R_0$ for different membrane models when 
$\eta=1.5$ and $1$. 
}
\label{fig4}
\end{figure}

In Fig.~\ref{fig4}(a), we plot $\Delta P_{\rm c}$ of the fluid membrane described by Eq.~(\ref{eqn:7a})
when $\eta=R_0/R^\ast=1.5$.
Naturally, $\Delta P_{\rm c}$ approaches to zero for $a/R_0 \to 1$.  
In Fig.~\ref{fig4}(b), we show the dependence of $\Delta P_{\rm c}$ on $1-a/R_0$ in the double 
logarithmic plot. The plot indicates the scaling relation 
$\Delta P_{\rm c}  \sim  ( 1 - a/R_0)^{\alpha}$.
The exponent is $\alpha=3/2$ for most cases, whereas the only exception is the case of $\eta=1$ 
(unswollen vesicle) for which we have $\alpha=7/2$.  

%%%%%%%%%%%%%%%%%%%%%%%%%%%
%%%%%%%%%%%%%%%%%%%%%%%%%%%

Further analytical  calculation of $\Delta P_{\rm c} $ is possible when $\varepsilon_a = 1- a/R_0$ is small.
For $\varepsilon_a \ll 1$, we obtain 
\begin{equation}
	\Delta P_{\rm c} 
	    \approx  \frac{16\sqrt{6}}{9} \frac{k_{\rm a}}{R_0}
	    \frac{\left[(\eta^2-1)\varepsilon_{a}+ (10-7\eta^2) \varepsilon_{a}^3/3 \right]^{3/2}}
	                {[(\eta^2-1) + 3(5 - 4\eta^2)\varepsilon_{a}^2]^{1/2}}.
                 \label{eqn:16}
\end{equation}
This expression explicitly shows that, apart from the special case of $\eta=1$, the exponent is $\alpha=3/2$.
In the limit of $\eta \gg 1$, the critical pressure becomes
\begin{equation}
\Delta P_{\rm c}\approx \frac{16\sqrt{6}}{9}\frac{\gamma}{R_0}\varepsilon_{a}^{3/2}, 
\end{equation}
where $\gamma= \eta^2 k_{\rm a} $ is the effective surface tension of the vesicle at the initial state. 
In the limit of $\eta \gg 1$, the area change of the vesicle during translocation is small, and the vesicle 
can be regarded as a liquid droplet (without any membrane) having the constant surface 
tension $\gamma$~\cite{deGennesbook}. Indeed, the same result can be obtained from the analysis of a liquid droplet whose
shape is controlled only by surface tension, as shown in Appendix B.

\subsection{Universality in the critical pressure}

We have shown that the critical pressure approaches zero as the
hole size approaches the equilibrium vesicle size according to the scaling relation
$\Delta P_{\rm c} \propto (1 -a/R_0)^{3/2}$.  
We now show that, apart from the exceptional case, the exponent $\alpha=3/2$ is 
always valid for spherical vesicles (for both fluid and elastic membranes) 
as long as the free energy is written in the form of Eq.~(\ref{eqn:1}). 

We start from the expression for the free energy in Eq.~(\ref{eqn:11}) 
that is valid for $\vert y \vert$ less than a certain value $y_{\rm m}>0$. This range of $y$ gets smaller as $\varepsilon_a$ gets smaller. 
Therefore we may expand  $A_{\rm tot}(y)$ with respect to $y$.   
Since  $A_{\rm tot}(y)$ is an even function of $y$, $G^*_y$ can be written as
\begin{equation}
G^*_y \approx - \Delta P V_0 y   - B_1 y^2 + B_2 y^4  + \cdots,         
\label{eqn:17}
\end{equation}
The coefficients $B_1$ and $B_2$ depend on the vesicle properties (e.g., size and elasticity) and 
also on the hole size $a$.  
For small $\Delta P>0$, $G^*_y$ has a local minimum at a certain value of $y$
that satisfies $-y_{\rm m} < y <0$.  
As $\Delta P$ increases, the local minimum disappears and $G^*_y$ becomes a monotonically 
decreasing function of $y$.
For $G^*_y$ to have such a shape change, both $B_1$ and $B_2$ must be positive, whereas the coefficient $B_1$ has to vanish as $\varepsilon_a \to 0$.  
Hence we may assume  $B_1= b_1 \varepsilon_a$ where $b_1$ is a positive constant.  
Then Eqs.~(\ref{eqn:9}) and ~(\ref{eqn:17}) give the following critical state
\begin{equation}
	y_{\rm c} =  -\left(\frac{b_1 \varepsilon_a }{6B_2}\right)^{1/2},  \quad   
	 \Delta P_{\rm c} = \frac{2\sqrt{6} b_1^{3/2} \varepsilon_a^{3/2}}{9 V_0 B_2^{1/2}}.                  
	 \label{eqn:19}
\end{equation}
The latter result shows that the relation $ \Delta P_{\rm c} \sim \varepsilon_a^{3/2} $ holds generally 
and is independent of the detail of the membrane property.

The singular behavior for a fluid membrane when $\eta=1$ (black line in Fig.~\ref{fig4}(b))
arises from the fact that $b_1$ in the above argument vanishes.
If the vesicle is swollen ($\eta>1$) or if the membrane is tethered, $b_1$ 
becomes nonzero and the exponent $\alpha=3/2$ is recovered.  
To demonstrate this, we calculate $\Delta P_{\rm c}$ for a rubber membrane with Eq.~(\ref{eqn:nh}) and the result is shown in Fig.~\ref{fig4} (b). 
For the rubber membrane, we see that the exponent $\alpha$ does not show any 
singularity at $\eta=1$, as shown by the orange and red lines in Fig.~\ref{fig4}(b).

\subsection{Universality in the translocation time}

Finally, we consider the vesicle translocation time $\tau$, i.e., the time needed for the vesicle to pass through the hole.   
It can be calculated by integrating $(dx/dt)^{-1}$ in Eq.~(\ref{eqn:7}) from the initial value 
to the final value:
\begin{equation}
	\tau = \int_{- x_{\rm m}}^{x_{\rm m}} dx \, \xi(x) \left( - \frac{dG^*_x}{dx} \right)^{-1},
       \label{eqn:20}
\end{equation}
where $x_{\rm m}$ defines the range of the integration.
The translocation time $\tau$ diverges as $\Delta P$ approaches to $\Delta P_{\rm c}$.  
The relation between $\tau$ and  $\Delta P - \Delta P_{\rm c}$ can be obtained by the similar 
phenomenological argument as before.

We consider the situation that $\Delta P $ is slightly larger than $\Delta P_{\rm c}$. 
We define a dimensionless parameter $\varepsilon_P =\Delta P/\Delta P_{\rm c} -1>0 $, which 
is small in the current situation.
Then, the translocation time is essentially determined by the time needed for the 
vesicle to pass through the small region near the critical point $x_{\rm c}$.  
In such a case, the friction coefficient $\xi(x)$ can be replaced by a constant $\xi_{\rm c} = \xi(x_{\rm c})$ 
and the derivative of $G^*_x$ can be expanded in terms of $x-x_{\rm c}$ as 
\begin{equation}
- \frac{dG^*_x}{dx} = C_0 + C_1(x-x_{\rm c}) + C_2(x-x_{\rm c})^2 + \cdots,
\label{translocation}
\end{equation}
where the coefficients $C_i$ ($i=0,1,2, \cdots$) are functions of $\varepsilon_P$. 
The condition for the critical state in Eq.~(\ref{eqn:8})  imposes that $C_0 $ and $C_1$ must vanish 
when $\varepsilon_P=0$.
We may thus assume $C_0= c_0 \varepsilon_P$ and  $C_1= c_1 \varepsilon_P$.  
In the limit of $\varepsilon_P \to 0$, the integral becomes independent of  $x_{\rm m}$ and
$c_1$, and approaches to the following asymptotic value:
\begin{equation}
	\tau \approx \int_{-\infty}^{\infty } dx \, 
	\frac{\xi_{\rm c}}{c_0 \varepsilon_P  + C_2 (x-x_{\rm c})^2 } 
	 = \frac{\pi \xi_{\rm c}\varepsilon_P^{-1/2}}{(c_0 C_2)^{1/2}} .	 
	 \label{eqn:22}
\end{equation}
Therefore, the translocation time is proportional to $(\Delta P-\Delta P_{\rm c})^{-1/2} $,
which is another universal property of the vesicle translocation.

\section{Summary and discussion}
In summary, we have shown that there is a universality in the translocation of a wide class
of vesicles which take spherical shapes in the free state:
(i) there is a critical pressure difference $\Delta P_{\rm c}$ for the translocation to take place,
(ii) the critical pressure obeys the scaling relation $\Delta P_{\rm c} \sim (R_0 - a)^{3/2}$,
and (iii) the translocation time diverges as $\tau \sim (\Delta P - \Delta P_{\rm c})^{-1/2}$. 
We have shown that they are in agreement with rigorous calculations for fluid membranes and 
an approximate calculation for rubber membranes.   

So far, there are scarce data (experimental or simulation) to be compared with our 
predictions.  
The critical pressure for a droplet passing through a channel was studied by numerically solving the  
hydrodynamic equations~\cite{Zhifeng}, and they are plotted by the red circles in Fig.~\ref{fig4}(a). 
Furthermore, we give a simple estimation of $\Delta P_{\rm c}$ by using Eq.~(\ref{eqn:16}).
In the experiment, the typical radius of a polymersome is $R_0 \approx 20$\,$\mu$m and its stretching 
modulus is $k_a \approx 0.1$\,N/m~\cite{Bohdana1999}. 
Considering $a/R_0=0.8$ and $\eta=2$, we obtain 
$\Delta P_{\rm c} \approx 3\times 10^3$\,Pa which is consistent with the above work~\cite{Zhifeng}.

Other work to be compared is the molecular dynamics simulation for the translocation 
of tethered vesicles by Ranguelov \textit{et al.}~\cite{Bogdan2022}.
They reported the existence of $\Delta P_{\rm c}$, and studied the exponent $\omega$ which describes 
the divergence of the translocation time as $\tau \sim (\Delta P - \Delta P_{\rm c})^{-\omega}$.  
They found that $\omega$ varies from 0.22 to 0.85 depending on  the hole size.
Although our theory is not in contradiction with their results, the reason for the deviation needs 
to be studied. 
More studies are needed to confirm the validity of the present theory.

%\appendix
\begin{acknowledgments}
B.Z., S.K., and M.D. acknowledge the startup fund of Wenzhou Institute, University of Chinese Academy 
of Sciences (Nos.\ WIUCASQD2022016, WIUCASQD2021041, and WIUCASQD2022004) 
and Oujiang Laboratory
(No.\ OJQDSP2022018).  
B.Z.\ and S.K.\ acknowledge the National Natural Science Foundation of China (NSFC) through 
Grants (Nos.\ 22203022, 12274098, and 12250710127).
\end{acknowledgments}

%%%%%%%%%%%%%%%%%%%%%%%%
\appendix
%%%%%%%%%%%%%%%%%%%%%%%%

\section{Proof of equivalence of the two ways in determining critical points}

We consider a free energy function $G(x,y,P)$ which has two independent variables $x$ and $y$ 
and include a parameter $P$.  
The equilibrium state is given by
\begin{equation}
\frac{\partial G}{\partial x}=0, \qquad  \frac{\partial G}{\partial y}=0.                      \label{eqn:a1}
\end{equation}
Suppose that the above local minimum becomes unstable at a certain critical value $P_{\rm c}$.
It can be obtained in two ways.
\begin{itemize}
	\item[(i)] $P_{\rm c}$ is obtained by solving the following set of equations
        \begin{align}
		&\frac{\partial G}{\partial x}=0, \qquad  \frac{\partial G}{\partial y}=0, \nonumber \\
		&\frac{\partial^2 G}{\partial x^2}\frac{\partial^2 G}{\partial y^2}-\left(\frac{\partial^2 G}{\partial x\partial y}\right)^2=0.   \label{eqn:a2}
		\end{align}
	The first two equations indicate that $G(x,y)$ becomes stationary at $(x_{\rm c},y_{\rm c})$, and the
	third equation indicates that the same stationary state is at the stability limit.
	\item[(ii)] We reduce the two variables function $G(x,y)$ to a one-variable function 
	\begin{equation}
		G^*_x= \mathop{{\rm min}}\limits_{y} G(x,y)  = G(x, y^*(x)),        \label{eqn:a3}
	\end{equation}
	where $y^*(x)$ is the solution of $\partial G/ \partial y=0$.  Then $P_{\rm c}$ is obtained by the 
	condition that the local minimum of $G^*_x$ becomes unstable at $P_{\rm c}$:
	\begin{equation}
		\frac{dG^*_x}{dx}= 0, \qquad  \frac{d^2G^*_x}{dx^2}= 0.     \label{eqn:a4}
	\end{equation}
\end{itemize}
These two methods are equivalent and give the same value for $P_{\rm c}$.  This is proven as follows.

We use the formula:
\begin{equation}
	\frac{d G(x,y^*(x))}{dx} = \frac{\partial G}{\partial x}
	+ \frac{\partial G}{\partial y} \frac{d y^*}{dx}.          \label{eqn:a5}
\end{equation}
Then using the relation  $\partial G(x,y^*)/\partial y=0$, we obtain   
\begin{equation}
	\frac{d G^*_x}{dx} = \frac{\partial G}{\partial x}
	+ \frac{\partial G}{\partial y} \frac{d y^*}{dx} = \frac{\partial G(x,y^*)}{\partial x}.   
	\label{eqn:a6}
\end{equation}
By using Eq.~(\ref{eqn:a5}), we have
\begin{equation}
	\frac{d^2 G^*_x}{dx^2}  = \frac{\partial^2 G}{\partial x^2} + \frac{\partial^2 G}{\partial x \partial y} \frac{d y^*}{dx} .  
	\label{eqn:a7}
\end{equation}
On the other hand, since $\partial G(x,y^*)/\partial y=0$, we have
\begin{equation}
	\frac{d}{dx}\left(\frac{\partial G(x,y^*)}{\partial y}\right) = \frac{\partial^2 G}{\partial y \partial x} + \frac{\partial^2 G}{\partial y^2} \frac{d y^*}{dx} =0.      	
\label{eqn:a8}
\end{equation}
Therefore, $d y^*/dx= -  (\partial^2 G/\partial x \partial y)/(\partial^2 G/\partial y^2)$.  
Hence the second condition in Eq.~(\ref{eqn:a4}) is written as
\begin{align}
	\frac{d^2 G^*_x}{dx^2}  & = \frac{\partial^2 G}{\partial x^2} - \frac{\left(\dfrac{\partial^2 G}{\partial x \partial y}\right)^2}{\dfrac{\partial^2 G}{\partial y^2}}  
\nonumber \\	
& =  \frac{  \dfrac{\partial^2 G}{\partial x^2}  \dfrac{\partial^2 G}{\partial y^2}   -\left(\dfrac{\partial^2 G}{\partial x \partial y}\right)^2}{ \dfrac{\partial^2 G}{\partial y^2}}=0.   	\label{eqn:a9}
\end{align}
This condition is equivalent to Eq.~(\ref{eqn:a2}). Similarly, we can prove the equivalence by using the variable $y$:
\begin{align}
	\frac{d^2 G^*_y}{dy^2}  & = \frac{\partial^2 G}{\partial y^2} - \frac{\left(\dfrac{\partial^2 G}{\partial x \partial y}\right)^2}{\dfrac{\partial^2 G}{\partial x^2}}  
\nonumber \\	
& =  \frac{  \dfrac{\partial^2 G}{\partial x^2}  \dfrac{\partial^2 G}{\partial y^2}   -\left(\dfrac{\partial^2 G}{\partial x \partial y}\right)^2}{ \dfrac{\partial^2 G}{\partial x^2}}=0.   	\label{eqn:a10}
\end{align}

%%%%%%%%%%%%%%%%%%%%%%%%%%
\section{Derivation of the liquid droplet case}

The free energy $G$ for the case of the liquid droplet surface has the following form:
\begin{align}
G(y)=\gamma S_1(y)+\gamma S_2(y)-\Delta P V_0 y,
\end{align}
where $y=V_2/V_0-1/2$ and $S_i(y)$ can be obtained from Eqs.~(\ref{eqgeometric1}) and (\ref{eqgeometric2}).
Minimizing $G(y)$ with respect to $y$, we obtain,
\begin{align}
	\Delta P(y, a,\eta)=\frac{3\gamma}{R_{\rm 0}} 
	\left(\frac{d S_1/S_0}{d y}+\frac{d S_2/S_0}{d y}\right).
\end{align}
By taking the same strategy as the fluid membrane case, the analytical expression of $\Delta P_{\rm c}$ is given as
\begin{align}
	\Delta P_{\rm c}\approx \frac{16\sqrt{6}\gamma \varepsilon_a^{3/2}}{9R_{\rm 0}}.
\end{align}

%%%%%%%%%%%%%%%%%%%%%%%%
%\newpage
%%%%%%%%%%%%%%%%

\clearpage

\end{document}